\documentstyle[12pt]{article}

\def\epem{e^+e^- \rightarrow \pi^+\pi^-}
\def\tpp{\tau^- \rightarrow \pi^-\pi^0 \nu_{\tau}}

\def\be{\begin{equation}}
\def\ee{\end{equation}}
\def\ro{\rho-\omega}
\begin{document}
\vspace*{.3cm}
\begin{flushright}
\large{CINVESTAV-FIS-19/95 \\ UCL-IPT-95-17}
\end{flushright}
\begin{center}
\LARGE{\bf S-matrix approach to two-pion production in $e^+e^-$
annihilation and $\tau$ decay}
 \end{center}
\vspace{.8cm}
\begin{center}
\Large A. Bernicha$^1$, G. L\'opez Castro$^2$ and J. Pestieau$^1$\\
%\medskip
\vspace*{.4cm}
{\normalsize{\it $^1$ Institut de Physique Th\'eorique, Universit\'e
 Catholique \\ \it de Louvain, B-1348 Louvain-la-Neuve, BELGIUM. \\
\it $^2$ Departamento de F\'\i sica, Cinvestav del IPN, Apdo. \\
\vspace{-.4cm} \it  Postal 14-740, 07000 M\'exico, D.F., MEXICO.}}
\vspace*{.4cm}
\end{center}

\thispagestyle{empty}
\centerline{ \bf Abstract}
\vspace{.3cm}
  Based on the S-matrix approach, we introduce a modified formula for the
$\pi^{\pm}$ electromagnetic form factor which describes very well the
experimental data in the energy region $2m_{\pi} \leq \sqrt{s} \leq
1.1$ GeV. Using the CVC hypothesis we predict $B(\tpp) = (24.75 \pm
0.38)\% $, in excellent agreement with recent experiments.

 \vspace{.5cm} PACS numbers:  13.35.Dx, 14.40.Cs, 11.30.Hv, 11.55.-m

\newpage
\setcounter{page}{1}
\vspace{2cm}

\begin{center}
\bf I. Introduction.
\end{center}

  The processes $\epem$ and $\tpp$ provide a clean environment for a
consistency check of the Conserved Vector Current (CVC) hypothesis [1].
Actually, the measurement of the $\pi^{\pm}$ electromagnetic form factor
in $e^+e^-$ annihilation is used to predict [2] the dominant hadronic decay
of the tau lepton, namely $\tpp$. The weak pion form factor involved in
$\tau$ decay is obtained by removing the (model-dependent)
I=0 contribution (arising from isospin violation and included
{\em via} $\ro$ mixing) from the measured pion electromagnetic form factor.

  In a previous paper [3] we have applied the S-matrix approach to the
$\epem$ data of Ref. [4] and determined the pole parameters of the
$\rho^0$ resonance. In particular, we have fitted the data of Ref.
[4] by assuming a
constant value for the strength of the $\ro$ mixing parameter and
using different parametrizations to account for the non-resonant
background. As a result, the pole position of the scattering amplitude
was found [3] to be insensitive to the specific background chosen to fit the
experimental data.

  The purpose of this Brief Report is two-fold. We first argue that the
pole position in $\epem$ is not modified by taking
the $\ro$ mixing parameter as a function of the center-of-mass
energy, as already suggested in recent papers [5]. Then we propose a new
parametrization for the scattering amplitude of $\epem$, based on the
S-matrix approach,
which looks very similar to the Breit-Wigner parametrization with an
energy-dependent width.
This results into an improvement in the quality of the fits (respect to
Ref. [3]) while the
pole position and $\ro$ mixing parameters remain unchanged (as it should
be). Finally, we make use of
CVC to predict the $\tpp$ branching ratio, which is found to be in excellent
agreement with recent experimental measurements.

\

\begin{center}
\bf II. Energy-dependent $\ro$ mixing.
\end{center}

  We start by giving a simple argument to show that the pole position
would not be changed if we choose the $\ro$ mixing parameter to be
$m^2_{\rho \omega}(s) \propto s$ (namely $m^2_{\rho-\omega}(0) = 0$), where
$\sqrt{s}$ is the total center-of-mass energy in $\epem$.

  Let us consider Eq.(7) of Ref. [3] and replace $y \rightarrow
y's/s_{\omega}\,^{\S}$\footnotetext{$^{\S}$In the Vector Meson Dominance
model, $y$ is related to the usual $\ro$ mixing strength through
$y=m^2_{\rho \omega}f_{\rho}/(m_{\rho}^2f_{\omega}) \simeq -2 \times
10^{-3}$ [3].}, where $s_V = m_V^2 -im_V\Gamma_V$.
 This yields the following expression for Eq. (7) of Ref. [3]:
 \begin{eqnarray}
F_{\pi}(s) &=& \frac{A}{s-s_{\rho}} \left ( 1 + \frac{y' s}{s_{\omega}}
\frac{m_{\omega}^2}{s-s_{\omega}} \right) + B(s) \nonumber \\
&=& \frac{A'}{s-s_{\rho}} \left ( 1 + y^{''}
\frac{m_{\omega}^2}{s-s_{\omega}} \right) + B(s),
\end{eqnarray}
where $A$ and $B(s)$ denote the residue at the pole and non-resonant
background terms, respectively. The second equality above follows from the
approximations: \begin{eqnarray*}
A' &\equiv& A \left ( 1 + y'\frac{m_{\omega}^2}{s_{\omega}}\right ) \approx
A(1+y'), \\
y^{''} &\equiv& \frac{y'}{1+ y'm_{\omega}^2/s_{\omega}} \approx
\frac{y'}{1+y'} \end{eqnarray*}
{\em i.e.} by neglecting small imaginary parts of order
$y'\Gamma_{\omega}/m_{\omega} \approx 10^{-5}$ [3].
  Thus, since introducing $m^2_{\rho\omega} \propto s$ is equivalent to a
redefinition of the residue at the pole and of the $\ro$ mixing
parameter, we conclude that the pole position would not be changed if we
take a constant or an energy-dependent $\ro$ mixing parameter.

\

\begin{center}
\bf III. Electromagnetic pion form factor.
\end{center}

  Next, we consider a new parametrization for the pion electromagnetic
form factor. This parametrization is obtained by modifying the pole term
in the following way:
\be
s-m_{\rho}^2 + im_{\rho}\Gamma_{\rho} \theta (\tilde{s}) \rightarrow  D(s)
\equiv
[1-ix(s)\theta (\tilde{s})] (s-m_{\rho}^2 + im_{\rho}\Gamma_{\rho} \theta
(\tilde{s})),
\ee
where $\theta (\tilde{s})$ is the step function, with argument $\tilde{s} =s -4
m_{\pi}^2$.

  Observe that if we chose:
\be
x(s) = -m_{\rho} \left ( \frac{\Gamma_{\rho}(s) -
\Gamma_{\rho}}{s-m_{\rho}^2} \right),
\ee
then Eq. (2) becomes:
\be
D(s) = s-m_{\rho}^2 + m_{\rho}\Gamma_{\rho} x(s)\theta (s-4m_{\pi}^2) +
im_{\rho} \Gamma_{\rho}(s)
\ee
which, when inserted in (1),  looks very similar to a Breit-Wigner with an
energy-dependent width, which we will chose to be:
\be
\Gamma_{\rho}(s) = \Gamma_{\rho} \left ( \frac{s-4m_{\pi}^2}{m_{\rho}^2
-4m_{\pi}^2} \right)^{3/2} \frac{m_{\rho}}{\sqrt{s}}\ \theta (s-4m_{\pi}^2)
\ee
with the obvious identification $\Gamma_{\rho} = \Gamma (m_{\rho}^2)$.

  Using Eq. (2) we are lead to modified
expressions for Eqs. (8), (9) and (15) of Ref. [3], namely:
\begin{eqnarray}
F_{\pi}^{(1)}(s) &=& \left ( -\ \frac{am_{\rho}^2}{D(s)} + b\right) \left( 1
+ \frac{ym_{\omega}^2}{s-s_{\omega}} \right) \\
F_{\pi}^{(2)}(s) &=&  -\ \frac{am_{\rho}^2}{D(s)}\left(  1
+ \frac{ym_{\omega}^2}{s-s_{\omega}} \right) + b\\
F_{\pi}^{(4)}(s) &=&  -\ \frac{am_{\rho}^2}{D(s)}\left(  1
+ \frac{ym_{\omega}^2}{s-s_{\omega}} \right)\left[ 1 +
b\left(\frac{s-m_{\rho}^2}{m_{\rho}^2} \right) \right]^{-1}.
\end{eqnarray}

  Using Eqs. (6-8), we have repeated the fits to the experimental data of
Barkov {\em et al.} [4] in the energy region $2m_{\pi} \leq \sqrt{s} \leq
1.1\ {\rm GeV}$. As in Ref. [3], the free parameters of the fit are
$m_{\rho},\ \Gamma_{\rho},\ a,\ b$ and $y$. The results of the best fits
are shown in Table 1.

   From a straightforward comparison of Table 1 and the
corresponding results in Ref. [3] (see particularly, Eqs. (10),
(11), (16) and Table I of that reference), we observe
that the quality of the fits are very similar. Furthermore, the pole
position, namely the numerical values of $m_{\rho}$ and $\Gamma_{\rho}$,
and of the $\ro$ mixing parameter $y$, are rather insensitive to the new
parametrizations (as it should be). The major effect of the new
parametrizations is observed in the numerical values of $a$ (the residue at
the pole) and $b$ (which describes the background).

An interesting consequence of the results in Table 1 is an
improvement in the value of $F_{\pi}(0)$, which should equal 1 (the charge
of $\pi^+$).
  Indeed, from Eqs. (6-8) and Table 1 we obtain:
\begin{eqnarray}
F_{\pi}^{(1)}(0) &=& a+b \nonumber \\
&=& 0.997 \pm 0.015\ (0.962 \pm 0.020) \nonumber \\
F_{\pi}^{(2)}(0) &=& a+b \nonumber \\
&=& 0.997 \pm 0.015\ (0.960 \pm 0.017) \\
F_{\pi}^{(4)}(0) &=& \frac{a}{1-b}  \nonumber \\
&=& 1.011 \pm 0.010\ (0.987 \pm 0.013) \nonumber
\end{eqnarray}
where the corresponding values obtained in Ref. [3] are shown in brackets.
An evident improvement is observed.

   Let us close the discussion on this new parametrization with a short
comment: using $F_{\pi}^{(4)}(s)$ (with imaginary parts and $y$ set to zero) we
are able to reproduce very well the data of Ref. [6] in the space-like
region $-0.253\ {\rm GeV}^2 \leq s \leq -0.015\ {\rm GeV}^2$.

\

\begin{center}
\bf IV. Prediction for $\tpp$.
\end{center}

  Finally, using the previous results on the pion electromagnetic form
factor, we consider the decay rate for $\tpp$.
 As is well known [2], the CVC hypothesis
allows to predict the decay rate for $\tau^- \rightarrow (2n\pi)^-
\nu_{\tau}$   in terms of the measured cross section
in $e^+e^- \rightarrow (2n\pi)^0$. Since for  the $\tpp$ case the
kinematical range extends up to $\sqrt{s}=m_{\tau}$,
let us point out that we have verified that our parametrizations
for $F_{\pi}(s)$ reproduce very well the data of $\epem$ in the energy
region from 1.1 GeV to $m_{\tau}$.

   The decay rate for $\tpp$ at the lowest order is given by [2]:
\begin{eqnarray}
\Gamma^0(\tpp) = \frac{G_F^2 |V_{ud}|^2m_{\tau}^3}{384 \pi^3}
\int_{4m_{\pi}^2}^{m_{\tau}^2} ds \hspace{-.5cm} && \left(
1+\frac{2s}{m_{\tau}^2}
\right) \left( 1-\frac{s}{m_{\tau}^2} \right)^2 \nonumber \\
 .\hspace{-.8cm}&&\left(\frac{
s-4m_{\pi}^2}{s} \right)^{3/2} |F_{\pi}^{I=1}(s)|^2
\end{eqnarray}
where $V_{ud}$ is the relevant Cabibbo-Kobayashi-Maskawa mixing angle. In
the above expression we have neglected isospin breaking
in the pion masses. The form
factor $F_{\pi}^{I=1}(s)$ in the Eq. (10) is obtained from Eqs.
(6-8) by removing the I=0 contribution due to $\ro$ mixing (namely, $y=0$).

  According to Ref. [7], after including the dominant
short-distance electroweak radiative corrections the expression for the
decay rate becomes:
\be
\Gamma(\tpp) = \left( 1 + \frac{2\alpha}{\pi} \ln \frac{M_Z}{m_{\tau}}
\right) \Gamma^0 (\tpp).
\ee
We have not included the effects of long-distance electromagnetic radiative
corrections, but we expect that they would not exceed 2.0 \% .

   In order to predict the branching ratio, we use Eqs. (6)-(8) with
$y=0$,
 the results of Table 1 and the following values of fundamental parameters
(ref. [7, 8]):
\begin{eqnarray*}
m_{\tau} &=& 1777.1 \pm 0.5 \ {\rm MeV} \\
G_F &=& 1.16639(2) \times 10^{-5}\ {\rm GeV}^{-2} \\
|V_{ud}| &=& 0.9750 \pm 0.0007.
\end{eqnarray*}

   With the above inputs we obtain:
\be
B(\tpp) = \left (\frac{\tau_{tau}}{2.956 \times
10^{-13} s} \right) \cdot \left\{ \begin{array}{ll}
(24.66 \pm 0.26)\% & \mbox{from Eq. (6)} \\
(24.62 \pm 0.26)\% & \mbox{from Eq. (7)} \\
(24.96 \pm 0.32)\% & \mbox{from Eq. (8)} \end{array} \right.
\ee
or, the simple average
\be
B(\tpp) = (24.75 \pm 0.38)\%
\ee
which is in excellent agreement with recent experimental measurements and
other theoretical calculations (see Table 2).
Eq. (13) includes the errors (added in quadrature) coming from
the fit to $\epem$ and the 1 \% error in the $\tau$ lifetime [8]:
$\tau_{\tau} = (295.6 \pm 3.1)\cdot 10^{-15}\ {\rm s}$.

  In summary, based on the S-matrix approach we have considered a
modified parametrization for the $\pi^{\pm}$ electromagnetic form factor,
which describes very well the experimental data of $\epem$ in the energy
region from threshold to 1.1 GeV. The pole position of the S-matrix
amplitude is not changed by this new parametrization. Using CVC, we have
predicted the $\tpp$ branching ratio, which is found to be in excellent
agreement with experiment.

\newpage
\begin{center}
TABLE CAPTIONS
\end{center}

\begin{enumerate}
\item Best fits to the pion electromagnetic form factor of
Ref. [4], using Eqs. (6-8).
\item Summary of recent experimental measurements (Exp.) and theoretical
results (Th.) for the $\tpp$
branching ratio. The errors in the first entry arise from use of $\epem$
data, the $\tau$ lifetime and radiative correction effects [9], respectively.
 \end{enumerate}
\newpage

\begin{table}[h]
\begin{tabular}{|c|c|c|c|c|c|c|}
\multicolumn{7}{c}{Table 1 } \\
\hline
 &$m_{\rho}$ (MeV) & $\Gamma_{\rho}$ (MeV) & $a$  & $b$ & $y (10^{-3})$
& $\chi/d.o.f$ \\
\hline\hline
$F_{\pi}^{(1)}$& $756.74\pm $ & $143.78 \pm $ & $1.236 \pm$ &
$-0.239 \pm$ & $-1.91 \pm$ & 0.998 \\ & $0.82$ & $1.16$ & $0.008$  &
$0.013$ & $0.15$ & \\
\hline
$F_{\pi}^{(2)}$& $756.58\pm $ & $144.05 \pm $ & $1.237 \pm$ &
$-0.240 \pm$ & $-1.91 \pm$ & 1.008 \\ & $0.82$ & $1.17$ & $0.008$  &
$0.013$ & $0.15$ & \\
\hline
$F_{\pi}^{(4)}$& $757.03\pm $ & $141.15 \pm $ & $1.206 \pm$ &
$-0.193 \pm$ & $-1.86 \pm$ & 0.899 \\ & $0.76$ & $1.18$ & $0.008$  &
$0.009$ & $0.15$ & \\
\hline
\end{tabular}
\end{table}

\

\begin{table}[h]
\begin{tabular}{|c|c|}
\multicolumn{2}{c}{Table 2} \\
\hline
Reference & $B(\tpp)$ (in \%) \\
\hline\hline
Th. [9] & $24.58 \pm 0.93 \pm 0.27 \pm 0.50$ \\
\hline
Th. [10] & $24.60 \pm 1.40$ \\
\hline
Th./Exp. [11] & $24.01 \pm 0.47$ \\
\hline
Exp. [8] & $25.20 \pm 0.40$ \\
\hline
Exp. [12] & $25.36 \pm 0.44$ \\
\hline
Exp. [13] & $25.78 \pm 0.64$ \\
\hline
\end{tabular}
\end{table}
\end{document}